%% file: main.tex
\documentclass[conference,9pt]{IEEEtran}

\input{defines.tex}

\IEEEoverridecommandlockouts
\usepackage{cite}
\usepackage{amsmath,amssymb,amsfonts}
\usepackage{algorithm}
\usepackage{algpseudocode}
\usepackage{graphicx}
\usepackage{subcaption}
\usepackage{textcomp}
\usepackage{xcolor} 
\usepackage{threeparttable} 
\usepackage{multirow}
\usepackage{enumitem}
\usepackage[export]{adjustbox}
\usepackage{tikz}
\usepackage{afterpage} 
\usepackage{color}
\usepackage{pifont}
\usepackage[T1]{fontenc}
\usepackage{array}
\usepackage{booktabs}
\usepackage{float}
\usepackage{stfloats}
\usepackage{hyperref}
\usepackage{caption}
\usepackage{listings}
\usepackage{fontawesome5}
\usepackage{notoccite}
\usepackage{arydshln}
\usepackage{vcell}
\usepackage{titlesec}

\titlespacing*{\section}{0pt}{1.0ex plus .2ex}{0.8ex plus .2ex}
\titlespacing*{\subsection}{0pt}{0.8ex plus .2ex}{0.6ex plus .2ex}
\usepackage[font=small]{caption}

\usepackage[margin=0.68in]{geometry}

\definecolor{codebgcolor}{rgb}{0.95,0.95,0.92}

\definecolor{rev}{rgb}{0,0,0}
\definecolor{rev2}{rgb}{0,0,0}
\definecolor{rev3}{rgb}{0,0,0}
\definecolor{rev4}{rgb}{0,0,0}
\def\ignorecitefornumbering#1{%
     \begingroup
         \@fileswfalse
         #1
    \endgroup
}
\makeatother
  
\def\BibTeX{{\rm B\kern-.05em{\sc i\kern-.025em b}\kern-.08em
    T\kern-.1667em\lower.7ex\hbox{E}\kern-.125emX}}

\begin{document}

\title{\vspace{-15pt}\textbf{\realprobetitle}}

\author{
    Jiho Kim, Cong Hao \\
    School of Electrical and Computer Engineering, Georgia Institute of Technology \\
    jiho.kim@gatech.edu, callie.hao@ece.gatech.edu
}

\maketitle

\input{./sections/abstract}
\begin{IEEEkeywords}
    FPGA, high-level synthesis, in-FPGA profiling
    \end{IEEEkeywords}

\input{./sections/intro}

\input{./sections/relatedwork}

\input{./sections/implementation}

\input{./sections/results}

\input{./sections/conclusion}
\input{./sections/acknowledgements}

\bibliographystyle{unsrt}

\bibliography{reference}

\end{document}

%% file: defines.tex
\newcommand{\tworows}[2]{\begin{tabular}[c]{@{}c@{}}#1\\#2\end{tabular}}

\newcommand{\myarraystretch}{1.3}

\newcommand\realprobetitle{\LARGE RealProbe: An Automated and Lightweight Performance Profiler for In-FPGA Execution of High-Level Synthesis Designs}

\usepackage{color}
\definecolor{highlighted}{rgb}{0.6,0.6,0.6}

\newcommand{\RPregLUTOH}{16.98\% }
\newcommand{\RPregFFOH}{43.15\% }

\newcommand{\benchmarks}{28 }

%% file: sections/abstract.tex
\begin{abstract}

High-level synthesis (HLS) accelerates FPGA design by rapidly generating diverse implementations using optimization directives.
However, even with cycle-accurate C/RTL co-simulation, the reported clock cycles often differ significantly from actual FPGA performance. This discrepancy hampers accurate bottleneck identification, leading to suboptimal design choices. Existing in-FPGA profiling tools, such as the Integrated Logic Analyzer (ILA), require tedious inspection of HLS-generated RTL and manual signal monitoring, reducing productivity.
To address these challenges, we introduce \texttt{RealProbe}, the first fully automated, lightweight in-FPGA profiling tool for HLS designs. With a single directive—\texttt{\#pragma HLS RealProbe}—the tool automatically generates all necessary code to profile cycle counts across the full function hierarchy, including submodules and loops. RealProbe extracts, records, and visualizes cycle counts with high precision, providing actionable insights into on-board performance.
RealProbe is non-intrusive, implemented as independent logic to ensure minimal impact on kernel functionality or timing. It also supports automated design space exploration (DSE), optimizing resource allocation based on FPGA constraints and module complexity. By leveraging incremental synthesis and implementation, DSE runs independently of the original HLS kernel.
Evaluated across \benchmarks{} diverse test cases, including a large-scale design, RealProbe achieves 100\% accuracy in capturing cycle counts with minimal logic overhead—just \RPregLUTOH{} LUTs, \RPregFFOH{} FFs, and 0\% BRAM usage. The tool, with full documentation and examples, is available on GitHub\footnote{\url{https://github.com/sharc-lab/RealProbe}}.

\end{abstract}

%% file: sections/intro.tex
\section{Introduction}\label{sec:introduction}

High-level synthesis (HLS) tools enable rapid FPGA design with high productivity. Vitis HLS~\cite{vitis-hls}, a representative tool, provides performance estimates through synthesis reports (C-synth) and cycle-accurate C/RTL co-simulation (Co-sim). However, \textbf{HLS performance estimates and simulations often deviate significantly from actual FPGA execution}. Fig.~\ref{fig_motivation} shows up to 103.8\% error in Co-sim cycle counts compared to in-FPGA results for two designs. These discrepancies arise because simulations cannot capture runtime dynamics such as memory controller behavior, bus contention, and DDR bandwidth, which introduce non-deterministic delays. As a result, identifying bottlenecks and optimizing designs becomes difficult.

In high-performance computing (HPC) and GPU domains, hardware-based profiling and visualization tools are essential for performance tuning~\cite{kousha2019gpu, agelastos2016hpc, sun2021daisen, zheng2012gmprof}. However, such tools are largely missing for HLS-based FPGA designs. Existing solutions offer limited visibility, require manual effort, or disrupt the original design. For example, Vitis Analyzer~\cite{vitis-analyzer} reports only top-level execution time, leaving submodules opaque. The Integrated Logic Analyzer (ILA)~\cite{ILA} demands manual Verilog-level instrumentation, reducing HLS productivity.
Among academic profiling tools, Curreri et al.~\cite{curreri2010performance} proposed an in-FPGA profiler using Impulse C, but it requires modifying source code. HLScope~\cite{choi2017hlscope} inserts FIFOs at the C++ level to analyze data stalls, but is intrusive, risks breaking functionality, and requires re-synthesis for each probe. These approaches lack flexibility and scalability for large or complex designs.

Despite these efforts, the FPGA community still lacks a profiling tool that is non-intrusive (no source changes), informative (beyond top-level), general (supporting varied analyses), and fully automated (no manual RTL inspection). Without such a tool, HLS developers are often left to \textit{guess} the causes of performance issues and rely on \textit{trial-and-error}, which significantly hinders productivity.

\input{./figs/fig_motiv.tex}

To address these limitations, we propose \texttt{RealProbe}, a pioneering automated tool designed to \textbf{measure actual in-FPGA execution performance of HLS designs}. Seamlessly integrated into the Vitis HLS and Vivado, \texttt{RealProbe} delivers comprehensive performance insights with zero manual effort. Its key features include:

\begin{itemize}[leftmargin=*]
\item \textbf{Comprehensive and Informative.}  
{RealProbe} profiles user-specified modules along with all internal submodules and loops across the entire design hierarchy, evaluated on designs with up to 285 submodules. It logs and visualizes results to provide detailed guidance for HLS performance analysis.

\item \textbf{Non-Intrusive and Decoupled.}  
Unlike previous tools, {RealProbe} does not modify the original HLS design, making it non-intrusive by preserving functionality and minimizing impact on timing (frequency) and physical implementation. Since it operates independently of the HLS kernel, changes to profiling targets only require incremental synthesis of {RealProbe}, leaving the original design untouched.

\item \textbf{Lightweight, Resource-Efficient, and Scalable.}  
{RealProbe} is optimized for minimal overhead, with zero BRAM usage and only 5.6\% runtime overhead. It automatically adapts to FPGA resource constraints and supports profiling of an arbitrary number of functions, subfunctions, and loops, handling up to $2^{64}$ cycles.

\item \textbf{Flexible Design Space Exploration}. RealProbe can automatically explore trade-offs between resource usage, DRAM bandwidth, and maximum frequency, via efficient incremental synthesis, helping users select the most suitable profiling setups.

\item \textbf{Accurate Profiling Results.}  
Experimental validation across \benchmarks{} test cases shows that {RealProbe} achieves 100\% accuracy, with results cross-verified using ILA.

\item \textbf{Fully Automated and Publicly Accessible.}  
{RealProbe} is fully automated from HLS source to in-FPGA deployment, integrated into Vitis HLS and Vivado without requiring environment setup, and the only user input is a single line \texttt{\#pragma HLS RealProbe}. The tool is publicly available with complete source code.
\end{itemize}

%% file: figs/fig_motiv.tex
\begin{figure}[t]
    \setlength{\belowcaptionskip}{-5pt}
    \includegraphics[trim={0.65cm 18cm 6cm 5.5cm},clip,width=3.55in]{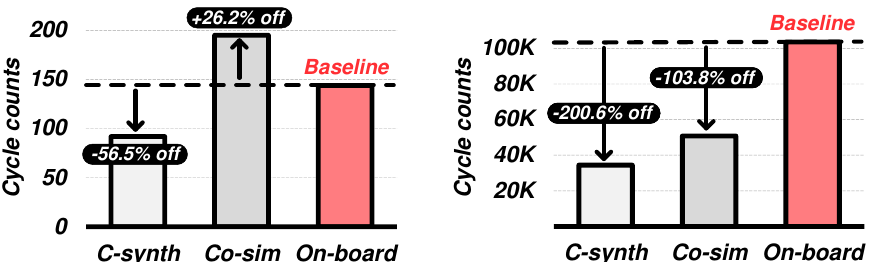}
    \caption{Cycle count discrepancies among HLS C-synth estimates, C/RTL co-simulation, and on-board execution for two HLS designs on the Pynq-Z2 FPGA. Left: array accumulation; right: matrix multiplication.}
    \label{fig_motivation}
    \vspace{-10pt}
\end{figure}

%% file: sections/relatedwork.tex
\section{Prior Work and Challenges}\label{sec:relatedwork}

\vspace{-4pt}
\subsection{Related work}
\vspace{-4pt}

\begin{figure}
    \centering
    \includegraphics[width=\linewidth]{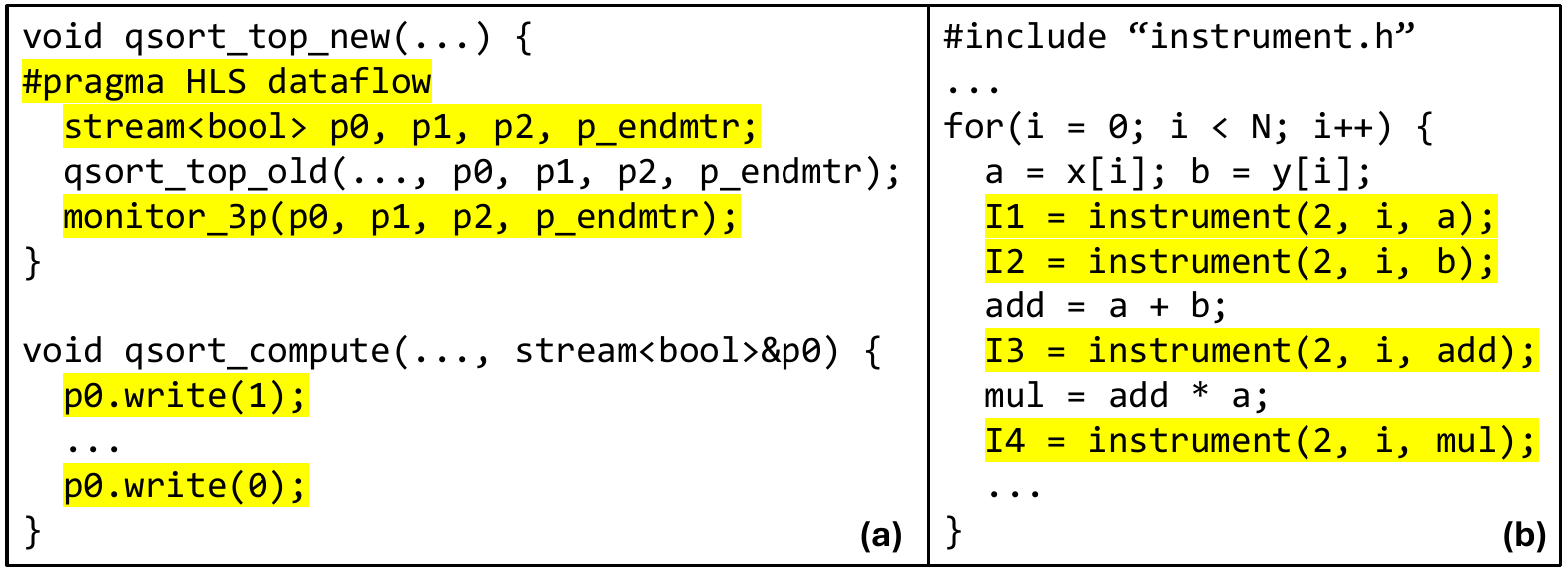}
    \caption{Previous \textbf{intrusive} profiling tools with highlighted instrumentation code. (a) HLScope~\cite{choi2017hlscope} requires manually inserting FIFOs and dataflow pragma. (b) Bensalem et al.~\cite{bensalem2020opencl} requires fine-grained instrumentation.}
    \label{fig:invasive}
\end{figure}

\input{./figs/fig_overview.tex} 

Various tools and techniques have been developed to support in-FPGA debugging and profiling, summarized in Table~\ref{tbl_comp}.
\textbf{Functional Debugging.}  
Goeders et al.~\cite{goeders2014effdebug} introduced a GDB-like framework that allows users to step through execution and inspect signals at C level. Calagar et al.~\cite{calagar2014srcdebug} developed a framework focused on functional verification. While effective for debugging functionality, these tools lack insight into performance characteristics.
\textbf{System-Level Profiling.}  
SoCLog~\cite{parnassos2016soclog} monitors system-wide activity, such as data movement at the bus level.
Pharos~\cite{rafii2021pharos} traces network-level events in multi-FPGA systems. Such tools provide only system-level performance metrics without fine-grained and internal analysis of modules and submodules.
\textbf{Intrusive or Semi-Automated Instrumentation.}  Several tools rely on intrusive instrumentation to collect performance data. Curreri et al.~\cite{curreri2010performance} proposed an RTL-level runtime analysis framework but not intended for HLS users. HLScope~\cite{choi2017hlscope} profiles performance by manually inserting FIFOs and dataflow pragmas into the HLS source. HLS\_print~\cite{sumeet2021hlsprint} and OpenCL-based tools~\cite{bensalem2020opencl} embed instrumentation directly into the source code.  
As illustrated in Figure~\ref{fig:invasive}, these tools require modifying the original HLS source, making them intrusive. Such approaches are not only tedious and error-prone, but also risk altering design functionality, introducing timing distortions, and ultimately yielding unreliable performance data.
\textbf{Others.}
While tools such as hls\_profiler~\cite{sumeet2022hls_profiler}, developed for Vitis HLS, support performance analysis based on signal activity from C/RTL co-simulation, they do not target actual hardware behavior and are not for on-board profiling.

\input{./tbls/tbl_comp.tex} 
\subsection{Challenges of In-FPGA Performance Profiling}\label{ssec:challenge}

Designing an in-FPGA profiler that operates directly on HLS designs in a non-intrusive manner presents several technical challenges. Below, we outline the key issues that must be addressed:

\begin{itemize}[leftmargin=*]

\item \textbf{C1: C++ and RTL signal alignment (Sec.~\ref{sec:address-C1}).}  
To enable performance profiling at the HLS source-code level, how can signals in the generated RTL be accurately identified and mapped back to their corresponding elements in the C++ source? This is particularly difficult since HLS tools often reorganize, inline, and optimize functions, loops, and statements during synthesis.

\item \textbf{C2: Probing deep hierarchies (Sec.~\ref{sec:address-C2}).}  
To ensure non-intrusive, how can deeply nested HLS components, such as submodules and loops, be instrumented without modifying the original design while keeping the instrumentation footprint minimal?

\item \textbf{C3: Minimal impact (Sec.~\ref{sec:address-C3}).}  
How can the profiler ensure minimal impact on the original HLS design, preserving critical metrics such as timing (maximum frequency) and maintaining the original implementation unchanged?

\item \textbf{C4: Resource efficiency and scalability (Sec.~\ref{sec:address-C4}).}  
How can the profiler scale to complex designs, supporting an arbitrary number of profiling targets and long execution traces, while minimizing resource overhead to ensure feasibility on real-world FPGAs?

\item \textbf{C5: Adaptive trade-off exploration (Sec.~\ref{sec:address-C5}).}
How can the profiler efficiently and automatically explore trade-offs among resource usage, clock frequency, and DRAM bandwidth, while adapting to a wide range of applications and target FPGA boards to identify the most suitable configuration?

\end{itemize}

%% file: figs/fig_overview.tex
\begin{figure*}[t]
    \setlength{\abovecaptionskip}{5pt}
    \setlength{\belowcaptionskip}{-10pt}
    \includegraphics[trim={0.1cm 21.8cm 17cm 0.1cm},clip,width=7.5in]{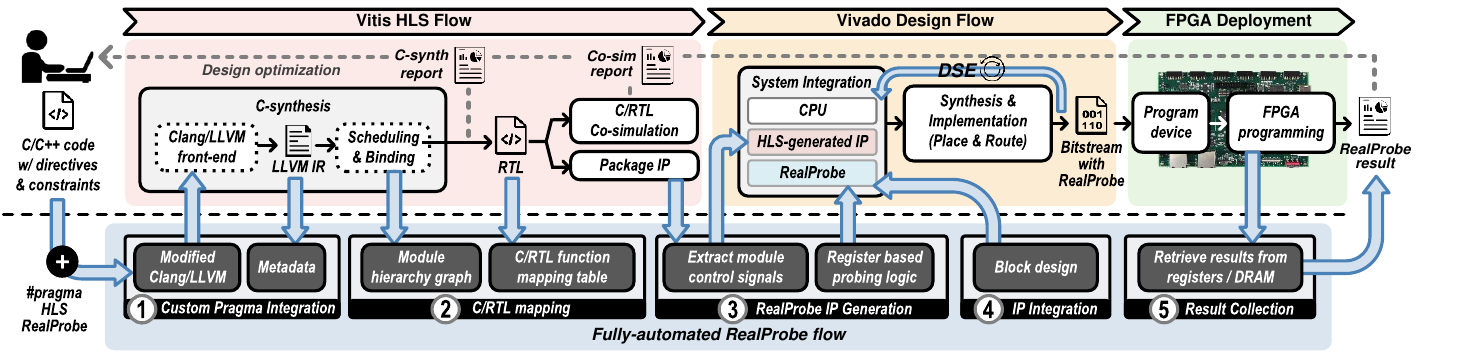}
    \caption{End-to-end automated RealProbe integrated with Vitis HLS and Vivado. The only user input for profiling is \texttt{\#pragma HLS RealProbe}.}
    \label{fig_overview}
\end{figure*}

%% file: tbls/tbl_comp.tex
\begin{table}[t]
    \centering
    \setlength{\belowcaptionskip}{0pt}
    \setlength{\tabcolsep}{1pt}
    \tiny
    \renewcommand{\arraystretch}{\myarraystretch}
    \begin{threeparttable}
    
    \caption{Comparisons of in-FPGA profiling tools for HLS designs.}
    \begin{tabular}{ccccccccccc} 
    \toprule

    \tworows{Feature}{}         & 
    \tworows{\textbf{RealProbe}}{[-1ex](this work)} & 
    \tworows{Geoders}{[-1ex]\cite{goeders2014effdebug}} & \tworows{Inspect}{[-1ex]\cite{calagar2014srcdebug}} & \tworows{SoCLog}{[-1ex]\cite{parnassos2016soclog}}  & \tworows{Curreri}{[-1ex]\cite{curreri2010performance}} & \tworows{HLScope}{[-1ex]\cite{choi2017hlscope}} & \tworows{HLS Print}{[-1ex]\cite{sumeet2021hlsprint}} & \tworows{OpenCL}{[-1ex]\cite{bensalem2020opencl}} & \tworows{CoChrono}{[-1ex]\cite{portas2024cochrono}} & \tworows{HPC}{[-1ex]\cite{huthmann2020extending}}\\ \hline

    Non-intrusive~\tnote{a}       & \faCheckCircle        & \faCheckCircle                                        & \faCheckCircle                                      & \faCheckCircle                                      & \faCheckCircle          & \textcolor{red}{\faTimes}    & \textcolor{red}{\faTimes} & \textcolor{red}{\faTimes} & \faCheckCircle         & \textcolor{red}{\faTimes}    \\\hline
    
    End-to-end automation~\tnote{b}        & \faCheckCircle        & \textcolor{red}{\faTimes}                             & \textcolor{red}{\faTimes}                           & \textcolor{red}{\faTimes}                           & \textcolor{red}{\faTimes} & \faCheckCircle             & \faCheckCircle          & \faCheckCircle          & \faCheckCircle         & \textcolor{red}{\faTimes}             \\\hline
    Profile internal hierarchy& \faCheckCircle        & \textcolor{red}{\faTimes}                             & $\triangle$                                         & \textcolor{red}{\faTimes}                           & $\triangle$         & $\triangle$             & $\triangle$          & $\triangle$          & \textcolor{red}{\faTimes}& $\triangle$             \\\hline

    C-RTL correlation~\tnote{c}& \faCheckCircle        & \textcolor{red}{\faTimes}                             & $\triangle$                                         & \textcolor{red}{\faTimes}                           & $\triangle$         & $\triangle$             & $\triangle$          & $\triangle$          & \textcolor{red}{\faTimes}& $\triangle$             \\\hline

    Cycle-level accurate                 & \faCheckCircle        & \textcolor{red}{\faTimes}                             & \textcolor{red}{\faTimes}                           & \faCheckCircle                                      & \faCheckCircle          & \faCheckCircle             & \textcolor{red}{\faTimes} & \faCheckCircle          & \textcolor{red}{\faTimes}& \faCheckCircle             \\\hline

    Fully-developed solution~\tnote{d}    & \faCheckCircle        & \faCheckCircle                                        & \faCheckCircle                                      & \faCheckCircle                                      & \textcolor{red}{\faTimes} & \textcolor{red}{\faTimes}    & \textcolor{red}{\faTimes} & \faCheckCircle          & \textcolor{red}{\faTimes}& \faCheckCircle             \\\hline
    Open source                 & \faCheckCircle        & \textcolor{red}{\faTimes}                             & \textcolor{red}{\faTimes}                           & \textcolor{red}{\faTimes}                           & \textcolor{red}{\faTimes} & \textcolor{red}{\faTimes}    & \textcolor{red}{\faTimes} & \textcolor{red}{\faTimes} & \textcolor{red}{\faTimes}& \textcolor{red}{\faTimes}    \\
    \bottomrule
     
    \end{tabular}\label{tbl_comp}
    \begin{tablenotes}
    \item [a]  Decoupled from the original HLS kernel; $^{\text{b}}$ Automated from HLS to in-FPGA execution; $^{\text{c}}$ Correlating profiled RTL signals back to original HLS; $^{\text{d}}$ Fully implemented and evaluated, rather than a conceptual or partial prototype.
    
    \end{tablenotes}
   
\end{threeparttable}
\vspace{-10pt}
    
\end{table}

%% file: sections/implementation.tex
\section{RealProbe Overview}
\label{sec:overview}

To address the challenges of low-overhead, non-intrusive performance profiling in HLS flows, we propose \textbf{RealProbe}—a fully automated, lightweight profiler for HLS designs. This section provides an overview of RealProbe’s five key stages, as illustrated in Fig.~\ref{fig_overview}:

\ding{202} \textbf{Custom pragma insertion}.  
To activate RealProbe, the user simply inserts the directive \texttt{\#pragma HLS RealProbe} into the C/C++ source code at the desired function or loop to be profiled. This pragma is interpreted by a modified Clang/LLVM front-end, generating metadata in the LLVM intermediate representation (IR) which links C/C++ constructs to the generated RTL constructs.

\ding{203} \textbf{Module extraction and C-to-RTL mapping}.  
Using the generated LLVM-IR metadata, RealProbe extracts the module hierarchy of the HLS kernel and constructs a hierarchy table to identify the relevant control signals to be externalized for profiling. RealProbe also constructs a C-to-RTL mapping table, associating C/C++ functions and loops with their corresponding RTL signals, enabling profiling data to be reported and analyzed directly at the source-code level.

\ding{204} \textbf{RealProbe IP generation}.  
Given the externalized signals, RealProbe generates a dedicated profiling IP block that connects to the HLS IP. A global clock counter, along with a set of performance counters, will log the timestamps when modules transition between active and idle states, capturing precise start and end cycles. To support long execution windows, the performance counter values are offloaded to DRAM when on-chip storage is full.
In addition, RealProbe supports automated design space exploration (DSE) to adapt resource usage based on FPGA constraints and profiling needs.

\ding{205} \textbf{System integration}.  
RealProbe automates the integration of its IP with the HLS IP in the Vivado block design flow. The RealProbe IP is synthesized and implemented independently from the HLS IP, so that it does not interfere with the functionality or performance of the original design. This decoupling also enables incremental synthesis—profiling a different function only requires updating the RealProbe IP, without re-synthesizing the HLS IP.

\ding{206} \textbf{Results collection}.  
RealProbe also automatically generates host code to launch the HLS kernel and retrieves profiling data. After kernel execution, the host reads the logged data from either on-chip buffers or DRAM. The collected data is mapped back to the original C/C++ code using the C-to-RTL mapping table. 
Fig.~\ref{fig_usage} illustrates an example HLS program instrumented by RealProbe to profile the \texttt{compute()} function, including its sub-functions \texttt{mult()} and \texttt{sum()}, and the \texttt{while} loop inside \texttt{sum()}. RealProbe reports execution iterations, total cycle counts, and start/end cycles for each sub-function, visualized as a waveform to highlight bottlenecks. Iterations after the fourth loop are omitted to reduce resource usage.

\input{./figs/fig_usage.tex}

\section{RealProbe Implementation}
\label{sec:implementation}

This section presents the detailed implementation of RealProbe and explains how it addresses the four challenges outlined in Sec.~\ref{ssec:challenge}.

\subsection{RealProbe Pragma and C-to-RTL Mapping -- Challenge C1}
\label{sec:address-C1}

The proposed directive, \texttt{\#pragma HLS RealProbe}, enables users to specify profiling targets without modifying the original source code. RealProbe leverages the Clang/LLVM front-end in Vitis HLS to maintain an accurate C-to-RTL mapping, ensuring robust associations even under aggressive compiler optimizations. In contrast, directly annotating C code is intrusive and unreliable, as HLS compilers determine which functions are synthesized and which signals are preserved. Optimizations like inlining and loop unrolling can obscure or eliminate annotations.

\underline{First}, to ensure consistency with existing HLS directives and enable profiling across entire module hierarchies, we extend the Vitis HLS front-end across multiple compiler phases. During lexical analysis, Clang is modified to recognize \texttt{\#pragma HLS RealProbe} and generate a unique token tagged with the pragma’s location in the source code. During semantic analysis, a dedicated \texttt{PragmaHandler} processes the pragma, associates it with the relevant function or loop, and injects metadata into the Abstract Syntax Tree (AST). During code generation, this metadata is propagated into the LLVM intermediate representation (IR). Finally, a custom backend pass is implemented that embeds the pragma metadata into the internal data structures of the Vitis HLS compilation flow. Positioned after LLVM IR generation, this pass attaches profiling information to the corresponding functions and loops, ensuring that the directives are preserved through scheduling and RTL generation. 

\underline{Second}, during HLS synthesis, RealProbe constructs a hierarchical tree representing the generated RTL modules. Each node in this tree corresponds to an RTL module, and its depth reflects its position within the hierarchy, as illustrated in Fig.~\ref{fig_mapping}. This tree serves as the basis for a one-to-one mapping table between C++ functions/loops and RTL signals. It also tracks submodule instances, their interconnections, and estimated cycle counts, which will guide resource budgeting---particularly the number of registers required in the RealProbe IP---which will be discussed in Sec.~\ref{sec:address-C3}.

An additional challenge arises from function inlining, which merges smaller functions into their callers and makes individual functions and loops unobservable. RealProbe addresses this by supporting customized inlining policies that override the default compiler behavior. This feature is discussed in more detail in Sec.~\ref{sec:inline}.

\subsection{Hierarchical Signal Extraction and Recording -- Challenge C2}
\label{sec:address-C2}

To profile components within deep module hierarchy, such as a \texttt{for} loop inside a submodule, RealProbe identifies and propagates control signals across all levels and collectively monitor them.

As shown in Fig.~\ref{fig_system}, HLS-generated IP uses finite-state machines (FSMs), driven by control signals such as \texttt{ap\_start} and \texttt{ap\_done}, to schedule operations across clock cycles. These signals reflect module and loop behavior (e.g., idle or active) and are used by the RealProbe IP to monitor the start and end of each component.

\underline{First}, RealProbe utilizes the hierarchy tree generated during synthesis to mark control signals at the lowest levels and propagate them upward through the module hierarchy. These signals are aggregated and routed to higher-level modules until they reach the top level. Once externalized, RealProbe updates the Vitis HLS IP’s XML file to declare them as external ports, specifying their names, directions, and data types. These ports are then connected to the RealProbe IP for signal monitoring.

\underline{Second}, to timestamp events across the design, RealProbe introduces a 32- or 64-bit \textbf{global cycle counter} that provides a unified timebase for all observations. Then, for each signal to be probed, a dedicated \textbf{performance counter} is created, which samples the global counter only on signal toggles. This edge-triggered sampling avoids redundant counters and reduces logic and register overhead. Performance counters can be implemented using shift register queues, BRAM blocks, or a combination of both, with configurable depths. These configurations can be tuned through DSE to balance on-chip resource usage and other overheads, as will be discussed in Sec.~\ref{sec:address-C5}.

Finally, during execution, whenever the shift registers or BRAMs used by performance counters become full, a \texttt{dump} signal is asserted to initiate the transfer of buffered timestamps to off-chip DRAM, reducing on-chip memory usage.

\input{./figs/fig_mapping.tex} 

\input{./figs/fig_system.tex} 

\subsection{Minimizing Impact on HLS Kernel -- Challenge C3}
\label{sec:address-C3}

The impact of RealProbe IP on the original HLS kernel must be minimized, including routing, timing behavior (maximum clock frequency), and synthesis runtime.
We propose two key techniques: (1) decoupled RealProbe IP integration for isolation, and (2) incremental synthesis and implementation to reduce runtime overhead, particularly when modifying profiling targets or adapting resource usage based on design space exploration (DSE) results.

\textbf{1. Decoupled RealProbe IP integration.}  
After extracting the signals to be probed, the RealProbe IP connects them to internal storage elements to record their active and idle states. As shown in Fig.~\ref{fig_system}, the RealProbe IP is instantiated as a \textit{standalone IP}, independent of the original HLS kernel.  
This decoupled approach provides two major benefits. First, it minimizes the impact on the original HLS design, with negligible effects on timing behavior and routability, as will be demonstrated in the experiments. Second, it enables incremental synthesis and implementation, allowing the RealProbe IP to be regenerated independently when profiling targets are updated, without touching the original HLS design.

\input{./figs/fig_incremental.tex}

\textbf{2. Incremental synthesis and implementation.}
A key advantage of being decoupled and non-intrusive is supporting efficient incremental changes. Existing profiling tools, including ILA, directly embed profiling logic into design source code~\cite{choi2017hlscope,curreri2010performance,bensalem2020opencl}, requiring full re-synthesis and re-implementation when profiling targets change, resulting in long turnaround times, especially for large designs.

RealProbe overcomes this limitation through incremental synthesis and implementation, as illustrated in Fig.~\ref{fig_incre}. During the initial run, it extracts all internal signals, including those deep in the hierarchy, and builds the C-to-RTL mapping table. This extraction is performed only once.
When users modify profiling targets (e.g., selecting a different module or submodule), RealProbe updates only the necessary connections and storage elements. Unused ports are deactivated using Vivado’s pull-down mechanism to maintain signal integrity. This process leverages Vivado’s incremental synthesis flow, preserving up to 99\% of cells, nets, and pins from the original implementation.

Notably, the incremental synthesis and implementation feature also offers another significant advantage: it enables RealProbe to automatically and efficiently adapt its resource allocation and perform DSE, which will be introduced later.

\subsection{RealProbe IP Resource Optimization -- Challenge C4}
\label{sec:address-C4}

To address Challenge C4, optimizing resource utilization while maintaining scalability, RealProbe employs several strategies to efficiently optimize on-chip resources, including registers, BRAMs, and LUTs.

\textbf{1. Initial resource allocation.}
RealProbe applies several initial optimizations to reduce on-chip resource usage for better scalability.
\underline{First}, performance counters for internal signals are allocated based on the number of modules to be profiled and estimated cycle counts from the HLS synthesis report. Since these estimates are often inaccurate, RealProbe conservatively doubles the projected requirements to ensure sufficient capacity. To prevent data loss, each counter includes a "full" signal that triggers offloading to DRAM when nearing capacity.
\underline{Second}, for loops, RealProbe captures only the first four iterations to minimize overhead. In pipelined loops, where the initiation interval (II) is fixed, all iterations behave identically, so that such truncation does not result in data loss. For non-pipelined loops with variable execution due to branching, iteration-level timing may vary. However, RealProbe still records the total loop execution time, ensuring accurate cycle counts.
\underline{Third}, to reduce LUT utilization, RealProbe optimizes its AXI read logic by partitioning the large monolithic multiplexer into smaller, hierarchy-based stages. This simple yet effective restructuring reduces LUT usage by an average of 58.0\%, as shown in Fig.~\ref{fig_resoptim}.

\input{figs/fig_resoptim}

\textbf{2. Resource allocation adaptation.}  
Supporting an arbitrary number of profiling modules is essential for a powerful profiler, but it can lead to excessive resource usage and potential implementation failures if FPGA capacity is exceeded. Existing tools such as ILA provide limited resource estimation, forcing users to rely on trial and error. To address this inefficiency, RealProbe introduces adaptive resource allocation both \textit{before} and \textit{after} implementation.

\underline{First}, prior to implementation, RealProbe assumes a conservative default configuration, for example, up to 50 modules with a maximum queue depth of 4, and uses only registers for the global cycle counter and performance counters to minimize BRAM consumption.
\underline{Second}, RealProbe applies an analytical model to estimate the resource usage of its profiling logic and extracts available resources from the C synthesis report. While these reports may not precisely match post-implementation usage, prior work has shown them to be conservatively estimated, making them a reasonable basis for initial planning~\cite{dai2018mlqor}. The LUT and FF usage of RealProbe is estimated as:
{\small
\begin{align*}
\text{LUT}_{\text{total}} &= C_{\text{axi}} + C_{\text{pc}} + C_{\text{decode}} \cdot \log_2(N) + \sum_{i=1}^{N} \left( C_{L1} + C_{L2} \cdot D_i \right), \\
\text{FF}_{\text{total}} &= C_{\text{axi}} + C_{\text{pc}} + C_{\text{decode}} \cdot \log_2(N) + \sum_{i=1}^{N} \left( C_{F1} + C_{F2} \cdot D_i \right),
\end{align*}
}
where $C_{\text{axi}}$ and $C_{\text{pc}}$ are fixed costs from the AXI interface and global counter logic, and $C_{\text{decode}}$ scales with the number of modules $N$. Each performance counter incurs a fixed cost ($C_{L1}, C_{F1}$) and a variable cost proportional to its depth $D_i$ ($C_{L2}, C_{F2}$).

\underline{Third}, if the estimated register usage exceeds availability, RealProbe checks BRAM resources and, if possible, replaces a portion of registers with BRAM-based storage. BRAM usage is also estimated analytically based on the number of modules and queue depths.

\underline{Fourth}, RealProbe proceeds with its first implementation and collects accurate post-implementation resource usage, including that of the original HLS kernel. It then refines its estimates of available resources, reapplies the analytical model, and adjusts the number of profiling modules and queue depths accordingly. Leveraging its support for incremental synthesis and implementation, RealProbe can iteratively adapt its resource allocation to fit within the FPGA, enabling scalable profiling for complex designs without risking implementation failure.

\input{figs/fig_resest}

\subsection{Automated Design Space Exploration -- Challenge C5}  
\label{sec:address-C5}

Another key challenge lies in balancing resource allocation: dedicating too many on-chip resources to profiling can negatively impact routability and timing closure (e.g., reduced maximum frequency, $F_{\text{max}}$), while allocating too few resources may require frequent offloading of profiling data to DRAM, potentially overwhelming memory bandwidth. Achieving the right balance is essential for both performance and scalability.
To navigate this trade-off, RealProbe performs automated DSE by considering three key metrics:

\begin{enumerate}[leftmargin=*]
    \item \textbf{Resource overhead} ($\Delta R_\text{util}$): The on-chip resources consumed by RealProbe, including registers, BRAMs, and LUTs, expressed as a weighted sum relative to the original design's resource usage:$\Delta R_\text{util} = \sum_{i} w_i \cdot \frac{R_{i,\text{RP}}}{R_{i,\text{origin}}}$,
    where $R_{i,\text{RP}}$ and $R_{i,\text{origin}}$ denote the usage of resource type $i$ by RealProbe and the original design, respectively. The weights $w_i$ can be adjusted to prioritize different resources, for example, favoring BRAM for larger profiling datasets or registers for smaller, low-footprint designs.

    \item \textbf{DRAM bandwidth overhead} ($\Delta B_\text{dram}$): The additional DRAM bandwidth required for offloading profiling data, calculated as a percentage relative to the original design’s DRAM bandwidth. The baseline DRAM bandwidth (\( B_\text{original} \)) is estimated using HLS C synthesis reports, based on memory interface specifications and burst configurations:$B_\text{original} = \frac{\text{Burst Size (Bytes)} \times \text{Total Number of Bursts}}{T_\text{total} \times T_\text{cycle}}$,
    where \( T_\text{total} \) is the total cycle count and \( T_\text{cycle} \) is the clock period.  

    To compute the overhead, RealProbe evaluates profiling configurations with different levels of offloading—e.g., 0\%, 25\%, 50\%, and 75\%—based on the total profiling data size \( S_\text{total} \), which depends on the number of modules \( N \) and profiling depth \( D_\text{depth} \). The DRAM traffic generated by profiling, \( S_\text{dram} \), is used to calculate the RealProbe DRAM bandwidth:$B_\text{with\_IP} = \frac{S_\text{dram}}{T_\text{total} \times T_\text{cycle}}$.

    \item \textbf{Maximum achievable frequency} ($F_\text{max}$): The peak timing performance of the design, which may degrade if added logic introduces routing congestion or extends critical paths. Both resource usage and DRAM interfacing can influence $F_\text{max}$, as reflected in post-implementation timing reports.
\end{enumerate}

\textit{Leveraging the incremental synthesis feature}, RealProbe automates the DSE process after the initial implementation. It explores configurations with varying resource allocation strategies (e.g., registers vs. BRAMs) and DRAM offloading levels (0\% to 75\%), using updated profiling targets and resource limits. Implementation results are fed into an automated feedback loop that constructs Pareto frontiers to visualize trade-offs between resource overhead, bandwidth usage, and timing performance.

\subsection{Handling Function Inlining}
\label{sec:inline}

Vitis HLS frequently applies function inlining as an optimization, embedding function bodies into their callers.
While this can reduce execution latency and enhance resource sharing, it complicates accurate C-to-RTL mapping, as illustrated in Fig.~\ref{fig_mapping}.
To enable flexible profiling without compromising the benefits of inlining, RealProbe extends inlining control by modifying the Clang/LLVM compilation flow. It provides three inlining options, allowing users to select the level of profiling granularity that best suits their design goals:

\begin{enumerate}[leftmargin=*]
    \item \texttt{Inline\_default}: Retains Vitis HLS’s default inlining behavior. If a profiled function is inlined, its behavior is attributed to the parent function during profiling.
    
    \item \texttt{Inline\_off\_all}: Disables automatic inlining for all functions, ensuring a one-to-one correspondence between C functions and RTL modules. This provides the most detailed and accurate profiling across the entire design.
    
    \item \texttt{Inline\_off\_top}: Disables inlining only for the top-level function specified by the \texttt{\#pragma HLS RealProbe} directive. This preserves inlining optimizations throughout the rest of the design while ensuring precise profiling for the targeted function.
\end{enumerate}

\subsection{System Integration, Automation, and Visualization}

The RealProbe IP is integrated into the Vivado block design, connecting seamlessly to the host, HLS-generated IP, and interconnect controllers. This integration is fully automated using a \textit{Tcl-only} workflow, requiring no manual intervention and preserving the original design syntax.  
Additionally, host code for controlling the RealProbe IP, collecting profiling results, and visualizing data is automatically generated as part of the workflow.

After deployment, users can initiate data collection via the \texttt{realprobe()} function. To avoid runtime overhead, RealProbe communicates with the host only after execution completes. Profiling results are displayed in tabular format or visualized using tools such as matplotlib, as shown in Fig.~\ref{fig_usage}. These visualizations provide detailed insights into execution behavior and enable comparison between co-simulation results and actual FPGA performance.

%% file: figs/fig_usage.tex
\begin{figure}
\vspace{-10pt}
    \setlength{\abovecaptionskip}{0pt}
    \setlength{\belowcaptionskip}{-5pt}
  \includegraphics[trim={0.22cm 24.4cm 16.5cm 0.1cm},clip,width=3.55in]{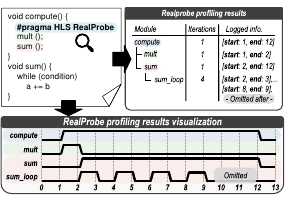}
  \caption{An example of using RealProbe pragma in an \textbf{non-intrusive} fashion, with the profiling results for function and loop hierarchy.}
  \label{fig_usage} 
  \vspace{-10pt}
\end{figure}

%% file: figs/fig_mapping.tex
\begin{figure}[t]
\vspace{-5pt}
    \setlength{\abovecaptionskip}{5pt}
    \setlength{\belowcaptionskip}{-10pt}
    \includegraphics[trim={0.2cm 19.9cm 7cm 0.5cm},clip,width=3.55in]{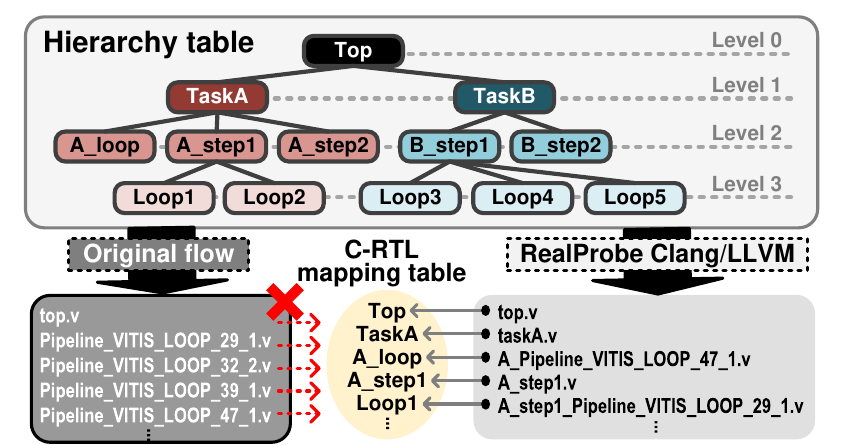}
    \caption{RealProbe’s modified LLVM flow accurately maps RTL modules to C functions, while the original flow fails (marked 'X').}
    \label{fig_mapping}
    
\end{figure}

%% file: figs/fig_system.tex
\begin{figure}[t]
    \setlength{\abovecaptionskip}{5pt}
    \setlength{\belowcaptionskip}{-15pt}
    \includegraphics[trim={0.65cm 19.4cm 7.5cm 0.5cm},clip,width=3.55in]{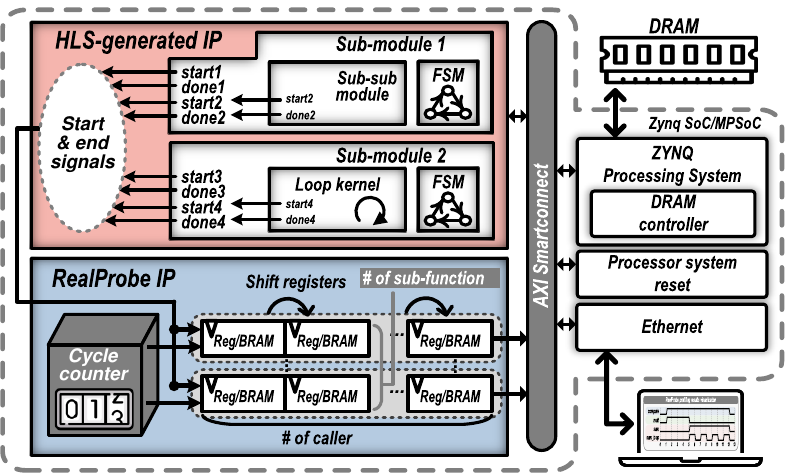}
    \caption{Integration with the HLS-generated IP and host: RealProbe IP captures control signals (e.g., \texttt{ap\_start}, \texttt{ap\_done}) across internal module hierarchies and loops, maintains internal counters, and writes counter values to DRAM as needed.}
    \label{fig_system}
\end{figure}

%% file: figs/fig_incremental.tex
\begin{figure}[t]
    \setlength{\abovecaptionskip}{5pt}
    \includegraphics[trim={0.3cm 23.5cm 8cm 0.3cm},clip,width=3.55in]{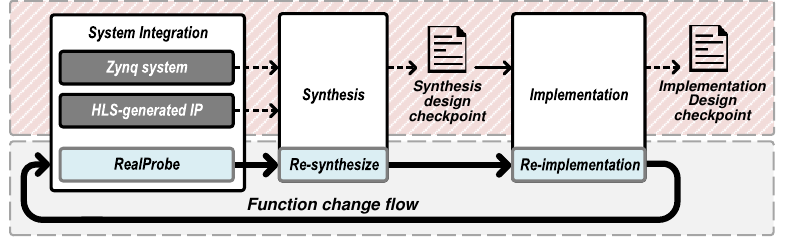}
    \caption{Incremental synthesis and implementation flow.}
    \label{fig_incre}
    \vspace{-6pt}
\end{figure}

%% file: figs/fig_resoptim.tex
\begin{figure}
    \setlength{\belowcaptionskip}{-12pt}
    \includegraphics[trim={0.1cm 0.1cm 0.1cm 0.1cm},clip,width=3.55in]{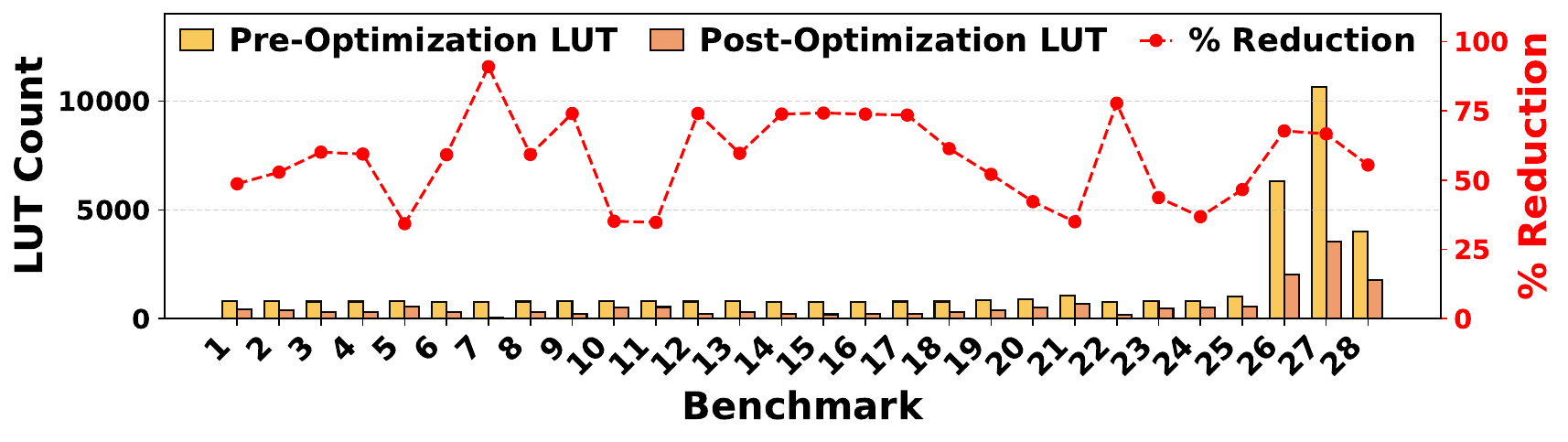}
    \caption{Effectiveness of RealProbe's LUT optimization.}
    \label{fig_resoptim} 
    \vspace{-5pt}
\end{figure}

%% file: figs/fig_resest.tex
\begin{figure}
    \setlength{\abovecaptionskip}{5pt}
    \setlength{\belowcaptionskip}{-12pt}
    \includegraphics[trim={0.1cm 0.1cm 0.1cm 0.1cm},clip,width=3.55in]{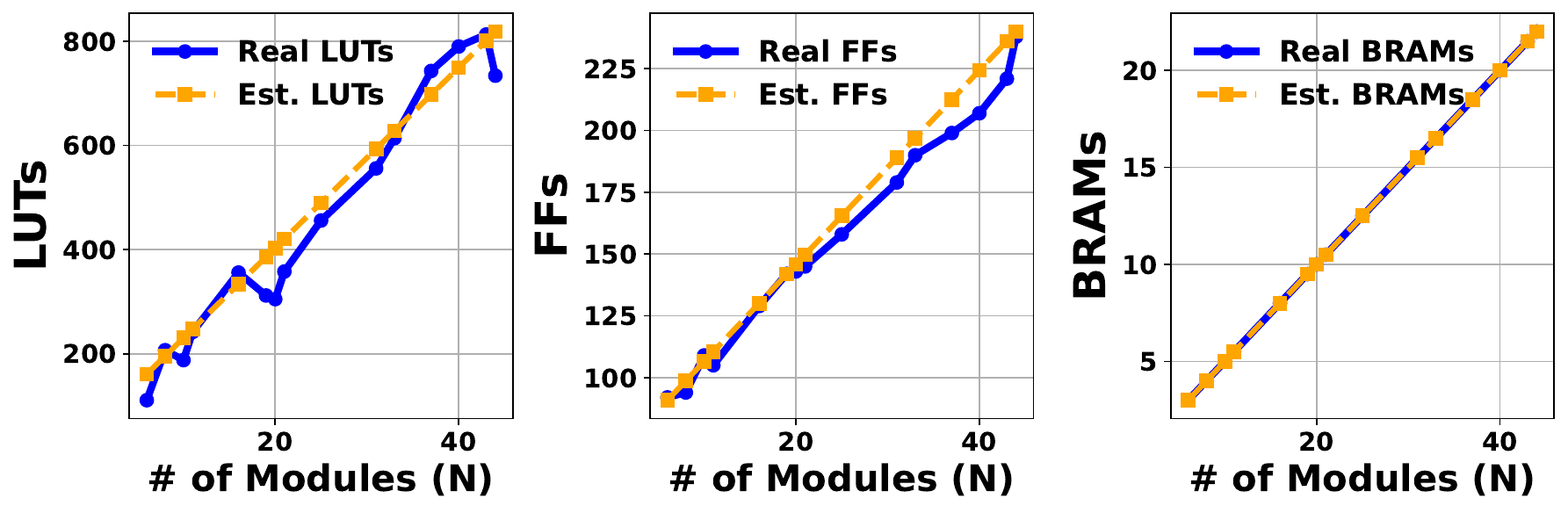}
    \caption{Analytical resource estimates v.s. post-implementation results.}
    \label{fig_resest} 
\end{figure}

%% file: sections/results.tex
\input{./tbls/tbl_experiments.tex}

\section{Experimental Results and Evaluation}

\subsection{Experimental Setup}
Experiments were conducted using Vitis HLS v2023.1.1, Vivado v2023.1.1, and the open-source Clang/LLVM frontend~\cite{vitisfrontend}, on a server running Red Hat Enterprise Linux 7.9 with a 64-core Intel Xeon Gold 6226R CPU. Vivado was configured with 64 parallel jobs for bitstream generation. Tests were performed on two FPGA platforms: the Pynq-Z2 (Zynq-7000 SoC) and the ZCU-102 (Zynq UltraScale+ MPSoC).

We evaluated \benchmarks HLS designs, including examples from Xilinx~\cite{vitis-example}, Kastner et al.~\cite{kastner2018parallel}, and five large applications on the ZCU-102: face detection, digit recognition (Rosetta~\cite{zhou2018rosetta}), and object detection (SkyNet~\cite{zhang2020skynet}).
To stress-test RealProbe’s scalability, we included two resource-intensive kernels: \textit{Kernel Selection} with 223 modules and \textit{SkyNet-big} with 285 modules, both utilizing over 90\% of FPGA resources. The number of tracked modules was initially capped at 50 and later adjusted as described in Section~\ref{sec:address-C3}.

\subsection{Profiling accuracy}
We first evaluate the profiling accuracy of RealProbe by comparing its results with those obtained manually using ILA. RealProbe and ILA are instantiated in the same system simultaneously, allowing them to observe the exact same execution. This co-instantiation eliminates potential discrepancies that can arise from system-level variability (e.g., DRAM access variations, host CPU interactions) when profiling tools are tested separately. 
We also confirm that all benchmarks produce the same functional outputs as the original HLS IP, whether RealProbe and ILA are present or not.

The profiled cycle counts are summarized in Table~\ref{tbl_experiments}, including C-synthesis and co-simulation results as comparisons. The number of modules indicates the complexity of the design. Since the cycle counts reported by ILA are exactly the same as RealProbe, we omit that column. The percentage value represent the differences between the cycle counts from ILA to C-synth, Co-sim, and RealProbe. 
To demonstrate RealProbe’s ability to profile internal modules with deep hierarchies, the right half of the table shows the bottleneck modules, which are sub-modules of the top-level module being profiled.
We analyze the results as follows:

\begin{itemize}[leftmargin=*]
    \item \textbf{RealProbe consistently matches the ILA results, with a 0.0\% error across all benchmarks and bottleneck modules}, demonstrating its high profiling accuracy without introducing errors.
    \item The cycle count difference between C-synth and ILA averages 37.8\%, with a striking discrepancy of 269.1\% for bottleneck modules. In some benchmarks, C-synth failed to provide program latency (showing as ? marks).
    
    \item Co-simulation results show a 34.0\% average difference from ILA cycle counts, and an even larger 383.0\% discrepancy for bottleneck modules, further proving that co-simulation alone is still insufficient to understand and optimize HLS design performance.
    
    \item ILA profiling is constrained by its internal buffer depth, as it uses a substantial amount of BRAM for signal storage, limiting its ability to track cycles beyond its maximum capacity of 131,072 cycles. This heavy reliance on BRAM restricts ILA to capturing only short segments of execution but impractical for long-running designs. For example, profiling the full execution of the digit recognition benchmark would require running the ILA over 5,541 times. In contrast, RealProbe efficiently utilizes BRAM for temporary storage and offloads profiling data to DRAM when registers or BRAM reach capacity, enabling continuous and scalable profiling without such restrictions.

\end{itemize}

\subsection{Impact of RealProbe: resource, runtime, timing, bandwidth}

\input{./figs/fig_resource.tex}

We evaluate the impact of RealProbe on the original HLS design, including resource overhead, runtime overhead, maximum frequency, and DRAM bandwidth overhead.

\textbf{1. Resource Overhead.}  
Fig.~\ref{fig_resource} compares the resource overhead of RealProbe and ILA relative to the original HLS designs. Since ILA’s resource usage depends on the number of tracked signals, we manually configured it to match RealProbe’s functionality for a fair comparison.
Overall, \textbf{RealProbe consistently incurs lower overhead than ILA, particularly in BRAM usage}. The average LUT overhead of RealProbe is \RPregLUTOH, ranging from 1.4\% to 53.5\%, compared to ILA’s average of 51.1\% (range: 15.8\%–188.7\%). For registers, RealProbe averages \RPregFFOH, with a range of 4.2\% to 144.8\%, while ILA averages 59.5\% (range: 16.4\%–247.4\%).
RealProbe also supports configurable storage using registers, BRAMs, or a hybrid of both. In contrast, ILA imposes significant BRAM overhead, averaging 1317.5\% of the original design’s BRAM usage, and reaching as high as 3650\%. This is because ILA is for fine-grained, signal-level debugging and is an overkill for profiling only.

\input{./figs/fig_runtime.tex}

\textbf{2. Synthesis Time Overhead.}  
As shown in Fig.~\ref{fig_tool_runtime}, RealProbe introduces an average synthesis runtime overhead of only 5.6\%, compared to 71.0\% for ILA. When profiling targets are updated, incremental synthesis significantly reduces this overhead: most designs reuse 99\% of cells, nets, and pins. On average, incremental synthesis takes just 9.1\% of the time required for full re-synthesis and re-implementation.

Notably, RealProbe also outperforms Co-sim in both runtime and practical usability. For example, the digit recognition benchmark required over 30 hours and 250GB of trace data to complete Co-simulation, whereas RealProbe completed hardware implementation and deployment in just 4 minutes and 12 seconds. In addition, RealProbe provides accurate, cycle-level performance breakdowns immediately after execution on hardware, making it a practical and efficient alternative to Co-sim, especially for large designs.

\textbf{3. Maximum Frequency Impact.}  
We evaluated the impact of RealProbe on maximum achievable frequency ($F_\text{max}$) using Vivado’s post-implementation timing analysis based on worst negative slack (WNS) at a target frequency. As shown in Table~\ref{tab:frequency}, RealProbe introduced minimal impact under two configurations, using registers only or BRAM only for performance counters, showing average increase (not decrease) of 1.74\% and 5.51\%, respectively.
This low impact is due to RealProbe’s non-intrusive design, where only control-flow signals are extracted as a standalone module, avoiding changes to the design’s core combinational and sequential logic.

\input{figs/fig_runtimedist}

\input{figs/fig_dse}

\input{./figs/fig_bottleneck.tex}

\input{tbls/tbl_frequency}

\textbf{4. DRAM Bandwidth Utilization} \label{sec:dram}  
One potential impact of RealProbe is its use of DRAM bandwidth, as profiling data may occasionally be offloaded to DRAM. Although RealProbe primarily relies on on-chip registers to minimize this usage, DRAM offloading can still occur in some scenarios. We evaluate this impact through both analytical modeling and empirical measurements.

Consider a case where a single module has a profiling depth of 64, with each entry using 64 bits, resulting in 512 bytes (0.5 KB) per dump. A dump occurs only after 64 status changes, i.e., idle/active transitions. Assume $N$ modules are being profiled at a 100 MHz clock, and each module toggles every $K$ cycles on average, then the worst-case DRAM bandwidth requirement is $
\frac{100{,}000{,}000}{64} \cdot \frac{1}{K} \cdot N \cdot 0.5~\text{KB} = 0.78 \cdot \frac{N}{K}~\text{GB/s}$.
For example, if $K=1000$ and $N=10$, the required bandwidth is only 0.0008 GB/s. Even under extreme (and unlikely) conditions ($K=1$), bandwidth usage would only be 0.8 GB/s. On the Pynq-Z2 board, we measured an effective DRAM bandwidth of 1.43 GB/s. In practice, higher-level modules toggle less frequently, further reducing bandwidth demands.

In addition, to quantitatively assess the runtime impact, we evaluated the Skynet and Kernel Selection benchmarks under four DRAM usage scenarios: 0\%, 25\%, 50\%, and 100\% offloading, as shown in Fig.~\ref{fig_rundist}. Skynet's runtime increased only slightly from 0.42 to 0.43 seconds, and Kernel Selection from 0.0359 to 0.0362 seconds. In a general matrix multiplication (GEMM) benchmark loading $512 \times 512$ blocks from DRAM, RealProbe introduced just 0.15\% overhead (231.11s to 231.45s). These results confirm RealProbe’s efficiency, maintaining low runtime overhead even under heavy DRAM access.

\subsection{DSE Results and Trade-offs}
We performed DSE to evaluate trade-offs among resource overhead, latency impact, and DRAM bandwidth under different RealProbe configurations. Fig.~\ref{fig_dse} shows results for Benchmarks 1, 24, and 25, comparing register-based (R) and BRAM-based (B) configurations under four DRAM dump ratios: 0\%, 25\%, 50\%, and 75\%.
Latency impact was derived from the measured total cycles and the maximum achievable frequency, then compared to the original design to compute relative latency overhead.

For Benchmark 1 with only 9 modules and shallow profiling depth, DRAM bandwidth overhead is negligible ($<0.1\%$). The register-based with 25\% DRAM dumping configuration (denoted by a diamond) offers the best trade-off between latency and resource usage.
For Benchmark 24 with 48 modules and deeper profiling requirements, we see benefits from BRAM-based setups, which better manage resource usage. The BRAM-based with 50\% DRAM dumping configuration offers balanced resource efficiency and latency, with DRAM overhead still negligible ($<0.08\%$). Latency remains under 3\% in most BRAM-based scenarios.
For Benchmark 25 with 26 modules, BRAM-based configurations again outperform register-based ones in terms of latency. DRAM usage remains minimal ($<0.1\%$), making latency and resource overhead the main optimization targets. The BRAM-based with 50\% DRAM dump ratio configuration offers the best overall balance.

\subsection{Results Visualization}
RealProbe enables users to visualize performance discrepancies between co-simulation and actual FPGA execution, uncovering true bottlenecks in HLS designs. Fig.~\ref{fig_bottleneck} demonstrates significant differences in profiling results from C-synthesis, co-simulation, and RealProbe. The timeline on the left provides a detailed breakdown of cycle counts, while the bump chart on the right illustrates the shifting bottleneck rankings across profiling stages. For example, C-synthesis identifies \texttt{B\_step2} as the primary bottleneck, co-simulation highlights \texttt{Loop1} and \texttt{Loop5}, yet RealProbe profiling on-FPGA identifies \texttt{Loop2} as the true bottleneck. These findings clearly indicate that relying solely on early-stage profiling methods can lead to misinterpretation and suboptimal optimizations, emphasizing the importance of accurate, real hardware-based profiling.

\subsection{Impact of Configurations on Co-sim and In-FPGA Discrepancies}

\label{:sec:why-different}

To investigate performance discrepancies between Co-sim and on-FPGA execution, we analyzed the \texttt{matrix multiplication} benchmark (Table~\ref{tbl_gemm}) under various configurations, including different FPGA platforms, optimization directives (e.g., pipelining, array partitioning), and implementation strategies (e.g., area, flow, and performance optimizations).

Our analysis reveals that optimization directives and implementation strategies have minimal effect on the discrepancies among C-synth, Co-sim, and RealProbe results. Instead, intensive DRAM access is the primary factor amplifying discrepancies between Co-sim and in-FPGA results, reaching up to 42.2\% on the Pynq-Z2 and 61.6\% on the ZCU-102. Notably, Co-sim reports identical cycle counts across both platforms.
This inaccuracy stems from Co-sim’s fixed empirical model for AXI read/write latencies, which fails to account for platform-specific DRAM characteristics and dynamic memory access patterns. For example, the Pynq-Z2 uses DDR3 (525 MHz, 512 MB), while the ZCU-102 features DDR4 (1200 MHz, 4 GB), leading to substantial differences in memory performance. These hardware-level factors are not captured by Co-sim, underscoring the importance of on-board profiling with RealProbe for accurate performance analysis.

\input{./tbls/tbl_gemm.tex}

%% file: tbls/tbl_experiments.tex
\begin{table*}[!bt]
    \centering
    \scriptsize
    \renewcommand{\arraystretch}{1.0}
    \setlength{\abovecaptionskip}{0pt}
    \setlength\tabcolsep{3pt}
    \begin{threeparttable}
    \caption{Clock cycle counts reported by C synthesis, C/RTL co-simulation waveforms, and RealProbe. Percentages in parentheses indicate differences relative to ILA results, which are omitted as they exactly match RealProbe values.}
    \label{tbl_experiments}
    \begin{tabular}{l|l|c|c:c:c|c:c:c}
    \hline
    \multirow{2}{*}{\textbf{No.}} & \multirow{2}{*}{\textbf{Benchmarks}} & \multirow{2}{*}{\tworows{\textbf{\# of}}{\textbf{modules}}} & \multicolumn{3}{c|}{\textbf{Total cycles}} & \multicolumn{3}{c}{\textbf{Bottleneck module cycles}} \\
    \cline{4-9}
    &&& \textbf{C-synth} & \textbf{Co-sim}~\tnote{a} & \textbf{RealProbe} & \textbf{C-synth} & \textbf{Co-sim}~\tnote{b} & \textbf{RealProbe} \\ 
    \hline
    \textcolor{rev}{1} & Fxp square root~\cite{vitis-example}                                & {9}   & 1580 (-14.9\%)       & 1700 (-6.8\%)         & 1816 ($\pm$0.0\%)         & 515 (-24.3\%)      & 560 (-17.6\%)    & 680 ($\pm$0.0\%)            \\
    \textcolor{rev}{2} & FIR filter~\cite{vitis-example}                                     & {8}   & 24121 (1.5\%)        & 23716 (-0.2\%)        & 23754 ($\pm$0.0\%)        & 254 (-21.8\%)      & 269 (-17.2\%)    & 325 ($\pm$0.0\%)            \\
    \textcolor{rev}{3} & Fxp conv~\cite{vitis-example}                                       & {7}   & 1565 (-14.7\%)       & 1692 (-6.1\%)         & 1795 ($\pm$0.0\%)         & 515 (-24.5\%)      & 562 (-17.6\%)    & 682 ($\pm$0.0\%)            \\
    \textcolor{rev}{4} & Fp conv~\cite{vitis-example}                                        & {7}   & 1568 (-14.7\%)       & 1692 (-6.3\%)         & 1799 ($\pm$0.0\%)         & 515 (-24.5\%)      & 562 (-17.6\%)    & 682 ($\pm$0.0\%)            \\
    \textcolor{rev}{5} & Parallel loops~\cite{vitis-example}                                 & {13}    & 159 (-27.0\%)        & 246 (+17.9\%)         & 202 ($\pm$0.0\%)          & 35 (-31.4\%)       & 36 (-29.4\%)     & 51 ($\pm$0.0\%)             \\
    \textcolor{rev}{6} & Imperfect loops~\cite{vitis-example}                                & {7}   & 473 (-7.4\%)         & 502 (-1.2\%)          & 508 ($\pm$0.0\%)          & 23 (-42.5\%)       & 24 (-40.0\%)       & 40 ($\pm$0.0\%)             \\
    \textcolor{rev}{7} & Max-bounded loops~\cite{vitis-example}                              & {2}   & 42 (-2.4\%)          & 71 (+39.4\%)           & 43 ($\pm$0.0\%)           & 42 (-2.3\%)        & 37 (-14.0\%)       & 43 ($\pm$0.0\%)             \\
    \textcolor{rev}{8} & Perfect nested loop~\cite{vitis-example}                            & {7}   & 473 (-6.8\%)         & 502 (-0.6\%)          & 505 ($\pm$0.0\%)          & 23 (-37.8\%)       & 24 (-35.1\%)     & 37 ($\pm$0.0\%)             \\
    \textcolor{rev}{9} & Pipelined nested loop~\cite{vitis-example}                          & {5}   & 441 (-2.9\%)         & 482 (+5.8\%)           & 454 ($\pm$0.0\%)          & 23 (-37.8\%)       & 24 (-35.1\%)     & 37 ($\pm$0.0\%)             \\
    \textcolor{rev}{10} & Static memory~\cite{vitis-example}                                  & {13}   & 117 (-9.4\%)         & 171 (+25.1\%)         & 128 ($\pm$0.0\%)          & 35 (-28.6\%)       & 36 (-26.5\%)     & 49 ($\pm$0.0\%)             \\
    \textcolor{rev}{11} & Pointer casting~\cite{vitis-example}                                & {13}   & 1452 (-1.0\%)        & 1499 (+2.2\%)         & 1466 ($\pm$0.0\%)         & 1042 (-0.5\%)      & 1042 (-0.5\%)    & 1047 ($\pm$0.0\%)           \\
    \textcolor{rev}{12} & Double pointer~\cite{vitis-example}                                 & {5}   & 29 (+10.3\%)         & 84 (+69.0\%)            & 26 ($\pm$0.0\%)           & 29 (+11.5\%)       & 26 ($\pm$0.0\%)        & 26 ($\pm$0.0\%)             \\
    \textcolor{rev}{13} & Multi-array access~\cite{vitis-example}                             & {7}   & 283 (-4.6\%)         & 322 (+8.1\%)          & 296 ($\pm$0.0\%)          & 131 (-9.7\%)       & 132 (-9.0\%)       & 145 ($\pm$0.0\%)            \\
    \textcolor{rev}{14} & Array access resolved~\cite{vitis-example}                          & {5}   & 275 (-4.7\%)         & 314 (+8.3\%)          & 288 ($\pm$0.0\%)          & 129 (+1.6\%)       & 257 (+102.4\%)   & 127 ($\pm$0.0\%)            \\
    \textcolor{rev}{15} & Fxp hamming window~\cite{vitis-example}                             & {5}   & 797 (-16.8\%)        & 868 (-7.3\%)           & 931 ($\pm$0.0\%)            & 259 (-21.5\%)      & 274 (-17.0\%)      & 330 ($\pm$0.0\%)      \\
    \textcolor{rev}{16} & Matrix multiplication~\cite{vitis-example}                          & {5}   & 34446 (-200.6\%)     & 50820 (-103.8\%)     & 103552 ($\pm$0.0\%)      & 34446 (-66.4\%)    & 50790 (-50.4\%) & 102403 ($\pm$0.0\%)        \\
    \textcolor{rev}{17} & Parallel merge loop~\cite{vitis-example}                            & {5}   & 143 (-214.0\%)       & 135 (-232.6\%)       & 449 ($\pm$0.0\%)          & 35 (-74.8\%)      & 89 (-36.0\%)    & 139 ($\pm$0.0\%)          \\
    \textcolor{rev}{18} & Access sequential~\cite{vitis-example}                              & {7}    & 91 (-48.4\%)         & 194 (+30.4\%)         & 135 ($\pm$0.0\%)          & 18 (+170.0\%)       & 32 (+3100.0\%)     & 1 ($\pm$0.0\%)              \\
    \textcolor{rev}{19} & Access w/ assert~\cite{vitis-example}                               & {21}    & 92 (-56.5\%)         & 195 (+26.2\%)         & 144 ($\pm$0.0\%)          & 18 (+800.0\%)        & 33 (+1550.0\%)     & 2 ($\pm$0.0\%)              \\
    \textcolor{rev}{20} & Access w/ dataflow~\cite{vitis-example}                             & {19}    & 32 (+96.9\%)         & 129 (+99.2\%)       & 11 ($\pm$0.0\%)              & 32 (+3100.0\%)           & 37 (+3600\%)         & 1 ($\pm$0.0\%)                \\
    \textcolor{rev}{21} & Unoptimized FFT~\cite{vitis-example}                                & {48}    & ? (?\%)              & 160603 (-0.2\%)         & 160861 ($\pm$0.0\%)      & 1027 (-25.9\%)     & 1138 (-17.9\%)   & 1386 ($\pm$0.0\%)          \\
    \textcolor{rev}{22} & Multi-stage FFT~\cite{vitis-example}                                & {4}    & ? (?\%)              & 5473 (-58.2.0\%)         & 8657 ($\pm$0.0\%)           & 1033 (-25.6\%)   & 1141 (-17.9\%)         & 1389 ($\pm$0.0\%)         \\
    \textcolor{rev}{23} & Huffman encoding~\cite{vitis-example}                               & {10}    & ? (?\%)              & 12150 (+27.5\%)        & 8812 ($\pm$0.0\%)              & 1318 (-75.7\%)  & 2256 (-58.3\%)        & 5416 ($\pm$0.0\%)         \\
    \textcolor{rev}{24} & Digit recognition~\cite{zhou2018rosetta}                            & {12}   & 144114007 (-96.9\%)   & 216100072 (-236.0\%)               &  726191018 (N/A\tnote{c} )              & 72010 (-80.2\%)      & 108004 (-70.2\%)    & 363029 ($\pm$0.0\%)        \\
    \textcolor{rev}{25} & Face detection~\cite{zhou2018rosetta}                               & {14}   & ? (?\%)               & 8030636 (-6.2\%)     &  8530151 ($\pm$0.0\%)              & ? (?\%)               & 2414577 (-65.2\%) & 6929825 ($\pm$0.0\%)          \\
    \textcolor{rev}{26} & Skynet~\cite{zhang2020skynet}                                       & {147}    & 22730587 (-15.8\%)     & 22724683 (-15.8\%)  &  26995099 (N/A\tnote{c} )          & 3457 (-87.9\%)      & 4531 (-84.1\%)    & 28541 ($\pm$0.0\%)         \\  
    \textcolor{rev}{27} & Skynet big~\cite{zhang2020skynet}                                   & {285}    & 19154068901 (-20.7\%)     & 19191631506 (-20.5\%)  &  24145455689 (N/A\tnote{c} )          &  51986(-52.2\%)      &  64510(-40.7\%)    & 108791 ($\pm$0.0\%)         \\  
    \textcolor{rev}{28} & Kernel selection                                                    & {223}    & ? (?\%)     & 264160 (-44.4\%)  &  475488 (N/A\tnote{c} )          & ? (?\%)      &  6840 (-30.3\%)    &  9817 ($\pm$0.0\%)         \\  
    
    \hline
    \end{tabular}
    \begin{tablenotes}
    \small
    \item [a] Cycle counts reported by C/RTL Co-simulation.  $^{\text{b}}$ Co-sim results for bottleneck module cycles are manually extracted from the waveform.
    \item [c] Cross-verification with ILA was not possible due to ILA's maximum depth limitation of 131,072 cycles.

    \end{tablenotes}
    \end{threeparttable}
    \vspace{-6pt}
    \end{table*}


%% file: figs/fig_resource.tex
\begin{figure}[t]
    \setlength{\abovecaptionskip}{0pt}
    \setlength{\belowcaptionskip}{-8pt}
    \includegraphics[trim={0.1cm 0.1cm 0.1cm 1cm},clip,width=3.55in]{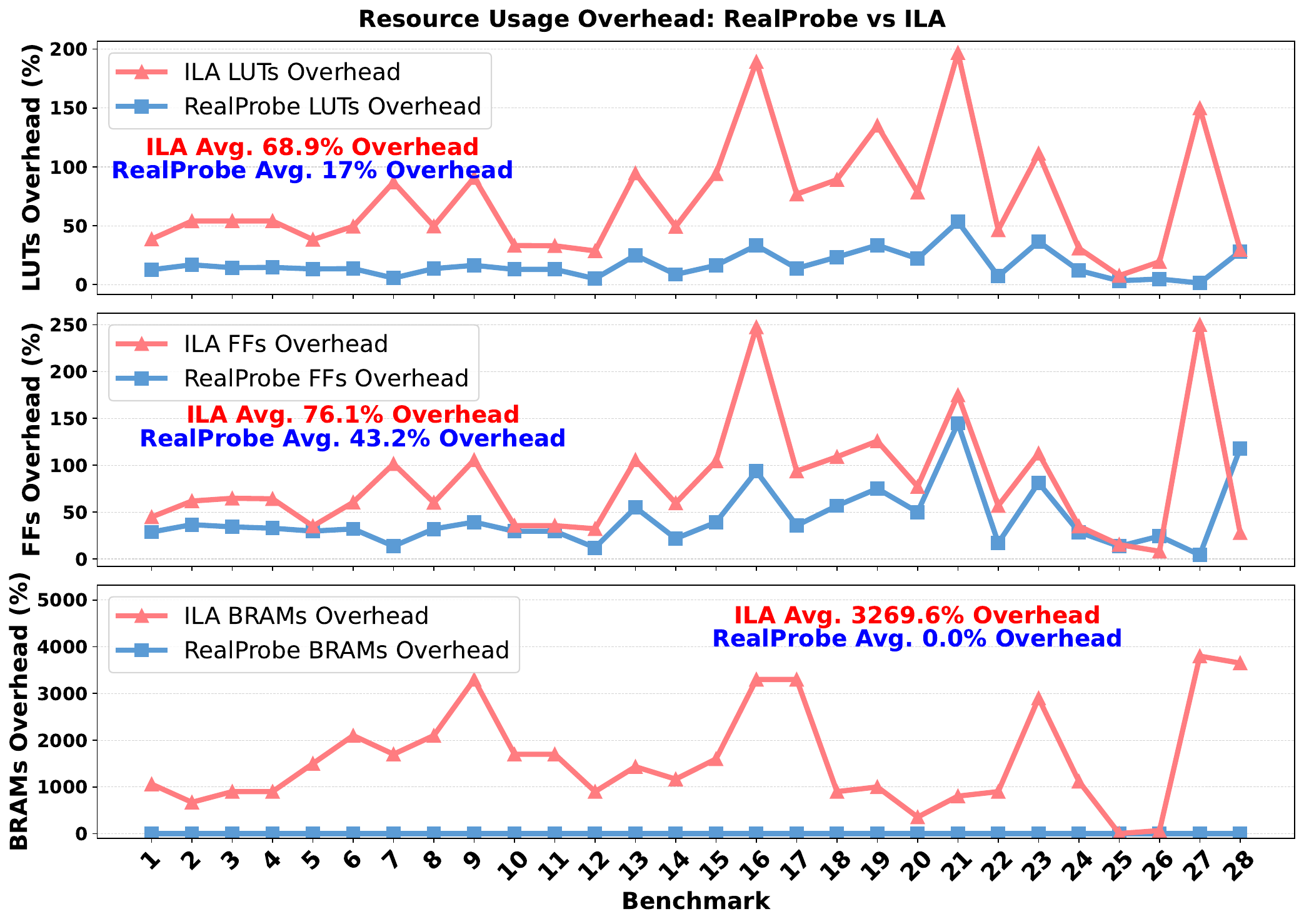}
    \caption{Resource overhead comparisons between RealProbe and ILA.}
    \label{fig_resource}
\end{figure}

%% file: figs/fig_runtime.tex
\begin{figure}
    \setlength{\abovecaptionskip}{0pt}
    \includegraphics[trim={0.1cm 0.1cm 0.1cm 0.1cm},clip,width=3.55in]{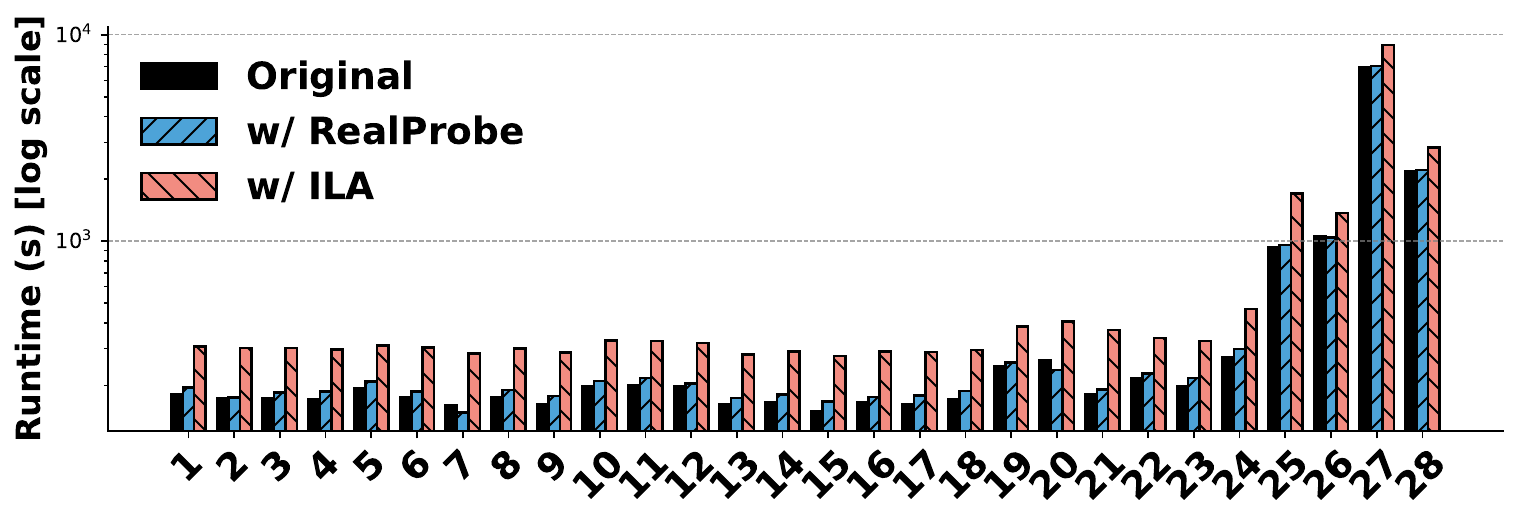}
    \caption{Runtime comparisons across the original design, with RealProbe, and with ILA. Since ILA can capture only a limited number of cycles, it requires multiple manual runs for some benchmarks. Here we only report the runtime for a single ILA execution at its maximum depth.}
    \label{fig_tool_runtime} 
\end{figure}

%% file: figs/fig_runtimedist.tex
\begin{figure}
    \includegraphics[trim={0.1cm 0.4cm 0.1cm 0.1cm},clip,width=3.55in]{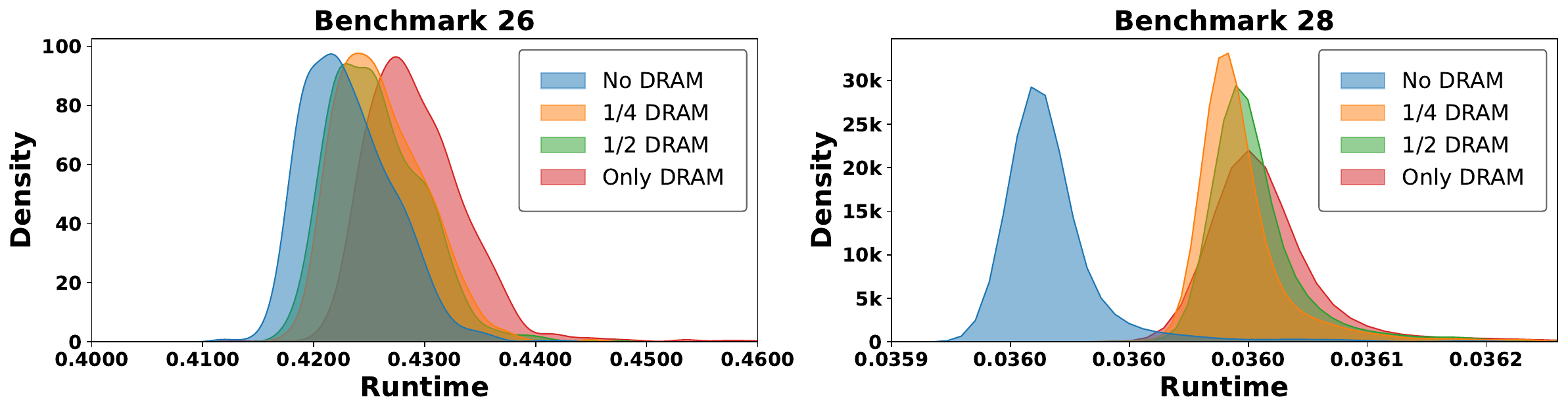}
    \caption{Runtime distribution by changing the amount of DRAM accesses for dumping recorded cycle values.}
    \label{fig_rundist} 
    \vspace{0pt}
\end{figure}

%% file: figs/fig_dse.tex
\begin{figure}[t]
    \includegraphics[trim={0.1cm 0.1cm 0.1cm 0.1cm},clip,width=3.55in]{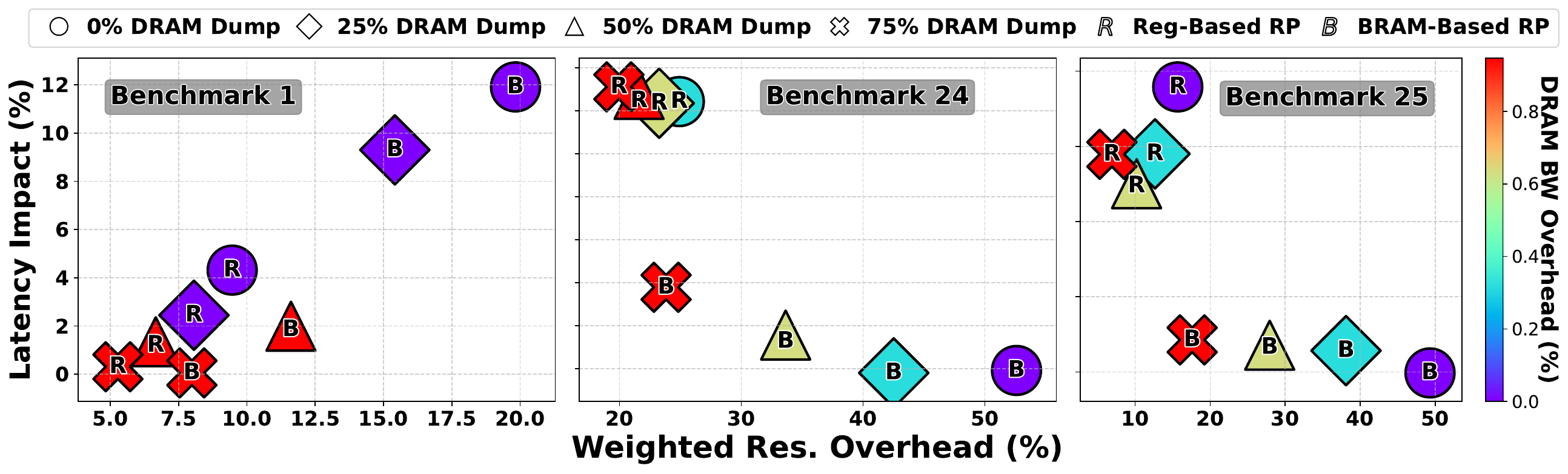}
    \caption{RealProbe DSE results, showing trade-offs across weighted resource overhead, latency impact, and DRAM bandwidth overhead. }
    \label{fig_dse} 
\end{figure}

%% file: figs/fig_bottleneck.tex
\begin{figure*}[t!]
    \setlength{\belowcaptionskip}{0pt}
    \includegraphics[trim={1.2cm 18.5cm 6cm 0.5cm},clip,width=7.2in]{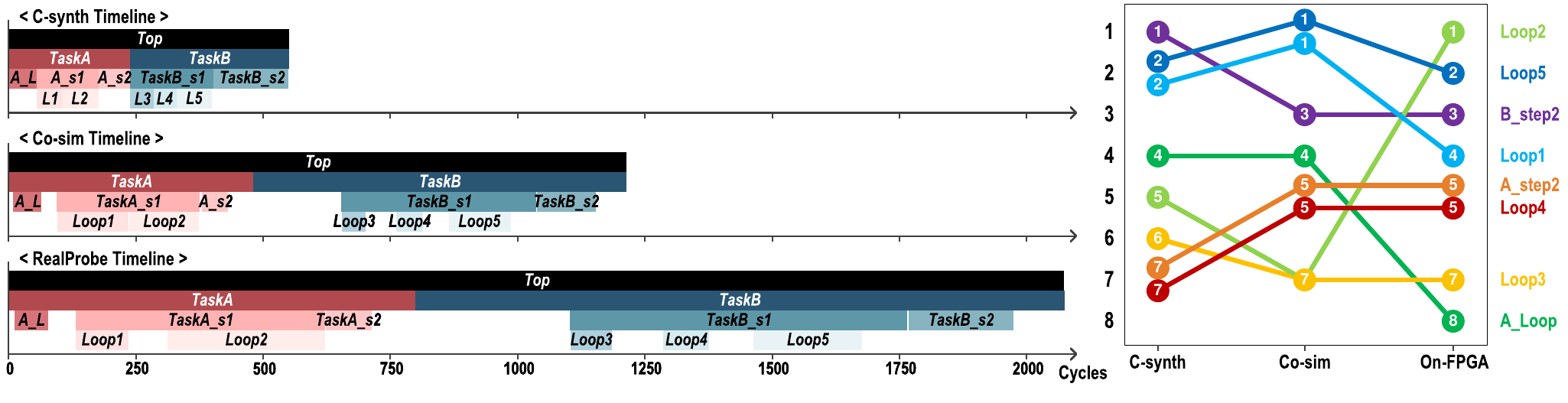}

    \caption{Execution timelines, cycle counts, and latency rankings vary across C-synthesis, co-simulation, and RealProbe profiling results. Discrepancies exist at the module, submodule, and loop levels; illustrated using the example used in Fig.~\ref{fig_mapping}.}
    \label{fig_bottleneck}
    \vspace{-4pt}
\end{figure*}


%% file: tbls/tbl_frequency.tex
\begin{table}[t]
\centering
\scriptsize
\renewcommand{\arraystretch}{0.95} 
\setlength{\tabcolsep}{9pt}
\caption{Maximum frequency impact of RealProbe variants.}
\begin{tabular}{c|c|cc|cc}
\hline

\multirow{2}{*}{\textbf{ID}} & \textbf{Original} 
& \multicolumn{2}{c|}{\textbf{Register only}} 
& \multicolumn{2}{c}{\textbf{BRAM only}} \\
\cline{3-6}
 & (MHz) & \textbf{Freq.} & \textbf{Change} & \textbf{Freq.} & \textbf{Change} \\

\hline
1 & 214.73 & 224.42 & +4.51\% & 243.78 & +13.53\% \\
2 & 241.90 & 233.21 & -3.59\% & 267.31 & +10.51\% \\
3 & 239.18 & 242.19 & +1.26\% & 249.69 & +4.39\% \\
4 & 243.19 & 238.21 & -2.05\% & 263.85 & +8.50\% \\
5 & 225.23 & 239.23 & +6.22\% & 248.88 & +10.50\% \\
6 & 212.54 & 218.96 & +3.02\% & 232.45 & +9.37\% \\
7 & 224.82 & 205.00 & -8.82\% & 216.68 & -3.62\% \\
8 & 260.82 & 244.08 & -6.42\% & 250.94 & -3.79\% \\
9 & 235.57 & 223.61 & -5.08\% & 244.56 & +3.82\% \\
10 & 231.05 & 223.26 & -3.37\% & 252.59 & +9.32\% \\
11 & 220.85 & 208.81 & -5.45\% & 235.90 & +6.82\% \\
12 & 227.95 & 227.74 & -0.09\% & 239.18 & +4.93\% \\
13 & 221.63 & 212.09 & -4.31\% & 217.49 & -1.87\% \\
14 & 227.63 & 234.85 & +3.17\% & 243.49 & +6.96\% \\
15 & 253.10 & 255.69 & +1.02\% & 256.67 & +1.41\% \\
16 & 305.16 & 296.65 & -2.79\% & 321.75 & +5.44\% \\
17 & 250.75 & 252.68 & +0.77\% & 263.44 & +5.06\% \\
18 & 247.89 & 237.64 & -4.13\% & 241.72 & -2.49\% \\
19 & 248.20 & 233.86 & -5.78\% & 233.81 & -5.80\% \\
20 & 195.27 & 193.31 & -1.01\% & 189.50 & -2.96\% \\
21 & 239.87 & 245.46 & +2.33\% & 249.31 & +3.94\% \\
22 & 249.69 & 238.89 & -4.32\% & 233.81 & -6.36\% \\
23 & 228.41 & 223.36 & -2.21\% & 248.26 & +8.69\% \\
24 & 207.34 & 223.46 & +7.78\% & 207.34 & 0.00\% \\
25 & 264.20 & 371.89 & +40.76\% & 293.34 & +11.03\% \\
26 & 207.73 & 240.50 & +15.78\% & 244.64 & +17.77\% \\
27 & 147.38 & 169.84 & +15.23\% & 189.11 & +28.31\% \\
28 & 155.71 & 165.32 & +6.17\% & 172.71 & +5.51\% \\
\hline
AVG & -- & -- & +1.74\% & -- & +5.51\% \\
\hline

\end{tabular}
\label{tab:frequency}
\vspace{-10pt}
\end{table}

%% file: tbls/tbl_gemm.tex
\begin{table}[!bt]
    \setlength{\tabcolsep}{8pt}
    \scriptsize
    \setlength{\abovecaptionskip}{5pt}
    \setlength{\belowcaptionskip}{5pt}
    \renewcommand{\arraystretch}{\myarraystretch}
    \begin{threeparttable}
    
    \caption{Total cycle count comparison of a matrix multiplication kernel with different configurations}
    \begin{tabular}{c|c|c|c} 
    \hline\hline
    \textbf{Variation}       & \textbf{C-synth}   & \textbf{Co-sim}    & \textbf{RealProbe} \\ \hline\hline

    Pynqz2 board (P) & 168025 (-0.3\%) & 168229 (-0.1\%) & 168449 \\\hline
    (P) w/ pipeline & 34847 (-1.2\%) & 35047 (-0.6\%) & 35271 \\\hline
    (P) w/ array part. & 34849 (-1.2\%) & 35047 (-0.6\%) & 35270 \\\hline
    (P) w/ unroll & 34849 (-1.2\%) & 35047 (-0.7\%) & 35281 \\\hline
    (P) w/ DRAM & 631873 (-65.9\%) & 1069174 (-42.2\%) & 1850604 \\\hline
    (P) w/ flow opt. & 168025 (-0.3\%) & 168229 (-0.1\%) & 168457 \\\hline
    (P) w/ perf opt. & 168025 (-0.3\%) & 168229 (-0.1\%) & 168460 \\\hline
    (P) w/ area opt. & 168025 (-0.3\%) & 168229 (-0.1\%) & 168448 \\\hline
    ZCU102 board (Z) & 69721 (-0.4\%) & 69922 (-0.1\%) & 69984 \\\hline
    (Z) w/ DRAM & 566337 (-79.7\%) & 1069174 (-61.6\%) & 2787078 \\\hline
    \hline
    \end{tabular}\label{tbl_gemm}
    
\end{threeparttable}
    \vspace{-15pt}
\end{table}

%% file: sections/conclusion.tex
\section{Conclusion}\label{sec:Conclusion}
This paper introduces RealProbe, the first fully automated and lightweight in-FPGA performance profiling tool for HLS. Seamlessly integrated into the Vitis HLS and Vivado flow, RealProbe uses a single pragma—\texttt{\#pragma HLS RealProbe}—to let users annotate functions for profiling. It then automatically extracts function and loop hierarchies, externalizes control signals, instantiates performance counters, and logs results to DRAM for host-side reporting.

RealProbe is non-intrusive, fully decoupled from the original HLS kernel, and scales to complex hierarchies. It adapts to on-chip resource constraints via incremental synthesis and supports automated design space exploration across resource usage, bandwidth, and timing. Fully integrated and publicly available, RealProbe significantly improves the usability and scalability of HLS performance profiling.

RealProbe incurs minimal in-FPGA overhead—\RPregLUTOH LUTs, \RPregFFOH FFs, 0\% BRAMs, and just 5.6\% runtime overhead—while achieving 100\% accurate cycle analysis compared to ILA. It has minimal impact on the maximum frequency and DRAM bandwidth of the original HLS kernel. Additionally, it enables side-by-side visualization of co-simulation and in-FPGA performance to help users identify and resolve design inefficiencies.

%% file: sections/acknowledgements.tex
\section*{Acknowledgement}\label{sec:Ack}
This research was supported in part by Semiconductor Research Corporation (SRC) and NSF (CSR-2317251 and CCF-2338365). 
This work would not have been possible without the generous support, insightful discussions, and invaluable guidance of Rishov Sarkar, Hanqiu Chen, and Stefan Abi-Karam. We also thank the anonymous reviewers for their thoughtful feedback and constructive suggestions.